\documentclass{article}
\usepackage[a4paper]{geometry}
\usepackage{amssymb}
\usepackage{euscript}
\usepackage{graphicx}
\usepackage{wrapfig}
\graphicspath{{images/}}
\usepackage{color}
\usepackage{epsfig}
\newcommand{\image}[3]{
\begin{figure}[#1]
\begin{center}
\includegraphics{full_#2.pdf}
\caption{\small#3}
\label{image:#2}
\end{center}
\end{figure}
}

\newcommand{\wrapimage}[4]{
\begin{wrapfigure}{#1}{#2}
\begin{center}
\includegraphics{full_#3.pdf}
\caption{\small#4}
\label{image:#3}
\end{center}
\end{wrapfigure}
}

\usepackage[colorlinks=true,hypertexnames=true]{hyperref}
\usepackage[style=nature,backend=bibtex,doi=false,isbn=false,url=true,alldates=year]{biblatex}

\def\sw{\sin\omega t}
\def\cw{\cos\omega t}

\begin{document}
\title{Using vibrating wire in non-linear regime as a thermometer in superfluid $^3$He-B}

\author{V.~V.~Zavjalov\/\thanks{e-mail: v.zavjalov@lancaster.ac.uk}}

\date{\today}
\maketitle

\begin{abstract}
Vibrating wires are common temperature probes in $^3$He experiments. By
measuring mechanical resonance of a wire driven by AC current in magnetic
field one can directly obtain temperature-dependent viscous damping. This
is easy to do in a linear regime where wire velocity is small enough
and damping force is proportional to velocity. At lowest temperatures in
superfluid $^3$He-B a strong non-linear damping appears and linear
regime shrinks to a very small velocity range. Expanding measurements to the
non-linear area can significantly improve sensitivity. In this note I
describe some technical details useful for analyzing such temperature
measurements.
\end{abstract}

\subsection*{Vibrating wire in the linear regime}

\wrapimage{r}{5.8cm}{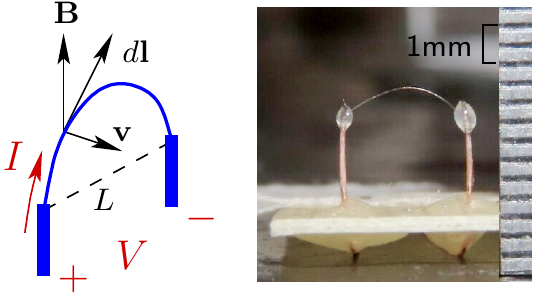}{Wire loop driven by current in magnetic field.
On the right image of a real device is shown (wire thickness $d=13\,\mu$m,
wire projection length $L=2.7\,$mm).}

First consider a wire loop with current~$I$ moving in magnetic field~$\bf B$
with velocity~${\bf v}\perp{\bf B}$. Geometry of this system is shown on Fig 1.
Force~$d{\bf F}$ acting on a piece of wire $d{\bf l}$ and EMF voltage~$dV$ across it are
\begin{equation}\label{eq:dFdV}
d{\bf F} = I\,[ d{\bf l} \times{\bf B}],
\qquad
dV= {\bf v}\cdot [d{\bf l}\times{\bf B}],
\end{equation}
where positive direction of current is along $d{\bf l}$ and positive voltage produces
current in the same direction (thus potential is decreasing along $d{\bf l}$).
By integrating along the wire one can find total force and voltage:
\begin{equation}
F = I L B,
\qquad
V= v L B,
\end{equation}
where $L$ is projection of the wire loop to a plane perpendicular to
magnetic field and $v$ is mean velocity along the wire length. Also
from~(\ref{eq:dFdV}) one can find mechanical power produced by the force:
\begin{equation}
W = \int {\bf v}\cdot d{\bf F} = I\,V.
\end{equation}

In the linear regime the wire driven by AC current can be represented as
a linear oscillator:
\begin{equation}
\ddot x = -\omega^2_0 x - \delta\dot x + \gamma\,\cw,
\qquad
\gamma=I L B/m_w
\end{equation}
where $x$ is average displacement, average velocity $v=\dot x$, and
parameters $m_w$, $\omega_0$, $\delta$ are effective mass of the wire,
resonance frequency, and resonance width. $\gamma$ is force divided by
mass $m_w$.

\subsection*{Non-linear regime in superfluid~$^3$He-B}

Behaviour of a vibrating wire in superfluid~$^3$He-B at low temperatures
was investigated long time ago in Lancaster~\cite{1989_fisher_wires,
1991_fisher_he3b_wire, 1995_enrico_wire_dumping}. There is a threshold
velocity at which wire can emit quasiparticles and dissipate extra
energy. This threshold is called "pair-breaking velocity", it is some
fraction of Landau velocity $\Delta/p_F$, usually 1/3 or smaller
depending on wire geometry. At zero pressure and low temperatures typical
value is less then 10~mm/s. Below the pair-breaking velocity damping is
determined by quasiparticle scattering, this is the non-linear regime we
are interested in. Force~$F_v$ acting on a wire moving with velocity $v$
has been derived by applying spectrum of Bogolubov quasiparticles to some
scattering model. There are a few results: for a simple 1D model, for
specular and diffusive 3D scattering. At the moment we are not interested
in exact expression, but there is an important result that the force can
be written via a temperature-independent function $\gamma_v$ of reduced
velocity~$v/v_0$:
\begin{equation}
F_v(\dot x) =
- m_w \delta_0 v_0 \gamma_v\left(\frac{v}{v_0}\right),
\end{equation}
where temperature-dependent parameters $\delta_0$ and $v_0$ are
\begin{equation}\label{eq:v0delta}
\delta_0 = \frac{p_F^2 v_F N(0)}{\rho_w\, d} \exp\left(-\frac{\Delta}{kT}\right),
\qquad
v_0 = \frac{kT}{p_F},
\end{equation}
$p_F$, $v_F$, $N(0)$ are $^3$He Fermi momentum, Fermi velocity, and
density of states at Fermi level, $\Delta$ is superfluid gap, $\rho_w$
and $d$ are density and diameter of the wire. Asymptotic behaviour of
function $\gamma_v$ at small velocity ($v\ll v_0$) should be linear,
which corresponds to the
linear regime: $F_v(\dot x) = - m_w\delta_0 v s_0$ with some temperature-independent dimensionless
calibration factor $s_0$. We also assume that function $\gamma_v$ is already
averaged along the wire loop and written for the average velocity~$v$.

Now the equation of motion is:
\begin{equation}
\ddot x = - \omega_0^2 x - \delta_i\dot x - \delta_0 v_0\gamma_v\left(\frac{\dot x}{v_0}\right) + \gamma\cw,
\end{equation}
where intrinsic damping of the wire is called $\delta_i$ and damping caused by interaction with helium
is described by the term with $\gamma_v$.

To find response at
frequency~$\omega$ we rewrite the equation of motion in van der Pol
coordinates:
\begin{equation}
u = x \cw - \frac{\dot x}{\omega} \sw,
\qquad
v = -x \sw - \frac{\dot x}{\omega} \cw.
\end{equation}
and average over period $0 < \omega t < 2\pi$. During this calculation we need
to average function $\gamma_v(\dot x/v_0)$ multiplied by
$\sw$ and $\cw$. This can be done by making substitution
\begin{equation}
u\,\omega/v_0 = c\cos\phi,
\qquad
v\,\omega/v_0 = c\sin\phi,
\end{equation}

shifting averaging period by $\phi$, and obtaining integrals with one parameter $c$:
\begin{eqnarray}
&&\langle \gamma_v(\dot x/v_0)\sw \rangle = \langle \gamma_v(- u\,\omega/v_0\sw - v\,\omega/v_0\cw)\sw \rangle =\\
&=&u\frac{\omega}{c v_0}\int_0^{2\pi} \gamma_v(-c\sw)\,\sw\,\frac{d\omega t}{2\pi} =
-\frac{u}{2}\frac{\omega}{v_0} S(c),
\end{eqnarray}
and similarly
\begin{equation}
\langle \gamma_v(\dot x/v_0)\cw \rangle = -\frac{v}{2}\frac{\omega}{v_0} S(c)
\end{equation}
where we introduce function $S(c)$ of positive dimensionless argument $c
= \frac{\omega}{v_0}\sqrt{u^2+v^2}$ as
\begin{equation}
S(c) = -\frac{2}{c}\int_0^{2\pi} \gamma_v(-c\sw)\,\sw\,\frac{d\omega t}{2\pi}
\end{equation}

\wrapimage{r}{5cm}{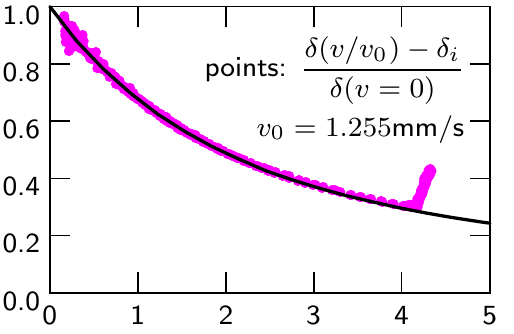}{Solid line: analytical function $S(c)$ for 1D
scattering model~(\ref{eq:bessel}). Points: an example of experimental data
scaled by choosing $v_0=1.255$~mm/s (see~Fig.~3 and (\ref{eq:delta}) for details).
Deviation at $c>4$ is the pair-breaking regime.
}

If function~$\gamma_v$ is known then function~$S$ can be calculated analytically
or numerically. Normally only a limited range of reduced velocities is
needed and $S$ can be easily tabulated or approximated by some smooth function.
For example, if damping is linear, $\gamma_v(x) = x$ then $S(c) = 1$.
For 1D scattering model~\cite{1989_fisher_wires} $S$ can be written via special functions:
\begin{eqnarray}
\gamma_v(x) &=& \mbox{sign}(x)(1-\exp(-|x|)),\\\label{eq:bessel}
S(c) &=& \frac{2}{c}\left(I_1(c) - L_{-1}(c) + \frac2{\pi}\right),
\end{eqnarray}
where $I_n(x)$ is modified Bessel function of first kind and $L_n(x)$ is
modified Struve function. For an arbitrary expansion
\begin{eqnarray}
\gamma_v(x) &=& a_1 x + \mbox{sign}(x)\,a_2 x^2 + a_3 x^3 + \mbox{sign}(x)\,a_4 x^3,\\
S(c) &=& a_1 + a_2 \frac{8c}{3\pi} + a_3 \frac{3c^2}{4} + a_4 \frac{32c^3}{15\pi}.
\end{eqnarray}
Note that $\gamma_v$ should be an odd function. In $^3$He-B it contains
quadratic term and have non-smooth second derivative at zero velocity.
This comes from the physical model where scattering channels open or
close when velocity sign changes. Quadratic term in $\gamma_x$ produces
linear term in $S(c)$. As it will be clear later this leads to a linear
shift of measured damping with driving force amplitude in AC
measurements and makes usual "linear" approach problematic.

On Fig.~2 analytical function $S(c)$ for 1D
scattering model~(\ref{eq:bessel}) is shown together with properly scaled
experimental data. One can see that shape of the function
is very close to the experiment. For practical purposes it is convenient to
approximate function $S(c)$ obtained in experiment with expression
\begin{equation}\label{eq:s0s1s2}
S(c) = \frac{s_0}{1 + s_1 c + s_2 c^2}
\end{equation}.

After averaging equation of motion, finding equilibrium $\langle \dot u
\rangle = \langle \dot v \rangle = 0$, and switching to a usual complex notation
we can write expression for complex velocity (bold font is used for complex values) as
\begin{equation}
{\bf v} = i\omega {\bf \gamma} \left(
\omega_0^2 - \omega^2 + i\delta_i\omega
  + i \delta_0\omega S\left(\frac{|{\bf v}|}{v_0}\right)
\right)^{-1},
\end{equation}
or in terms of current and voltage:
\begin{equation}\label{eq:nonlinVf}
{\bf V} = {\bf I} \frac{(LB)^2}{m_w}\,i\omega\left(
\omega_0^2 - \omega^2 + i\omega\delta_i
  + i\omega\delta_0S\left(\frac{|{\bf V}|}{v_0 LB}\right)
\right)^{-1}.
\end{equation}
This is a non-linear analog of Lorentzian formula which can be used for
fitting response of a vibrating wire. Voltage (or velocity) enters both
sides of the expression, but is can be calculated iteratively by starting
with ${\bf V}=0$ and calculating next-step value using the previous one
in the right-hand side.

\subsection*{Obtaining function $S(c)$ in experiment}

Consider a vibrating wire connected to an AC current source and a lock-in
amplifier. First a "frequency sweep" is done: voltage components are
measured as a function of frequency at a constant current amplitude. If
amplitude is small and the wire is close to the linear regime, then data can be
fitted with Lorentzian curve using a few parameters: complex amplitude,
resonance frequency, damping and a complex offset. If we are not in the
linear regime then after measuring function $S$ the fit can be corrected
by using nonlinear expression~(\ref{eq:nonlinVf}) to iteratively improve
the analysis. As a result we have a function
\begin{equation}
{\bf V} = \frac{{\bf A_0}\,i\omega}{\omega_0^2 - \omega^2 + i\delta\omega} + {\bf V_0}.
\end{equation}
with complex parameters ${\bf A_0}$ and ${\bf V_0}$, and real parameters
$\omega_0$ and $\delta$. We can not separate intrinsic dumping here and
use some total effective damping $\delta$ which is not very important for
the following analysis.

Voltage offset always exists because of parasitic coupling in the electric
circuit. Usually it is proportional to the drive current and weakly
depends on frequency. It is reasonable to measure offset once and then
subtract it from all data. When fitting the frequency sweep an offset
parameter~${\bf V_0}$ is still needed to compensate some time-dependent
drifts and non-linearities in the circuit, but in this case just a
constant offset (without any frequency dependence) can be used here.

When we know ${\bf A_0}$ and ${\bf V_0}$ parameters, we can do an
"amplitude sweep": set generator frequency close to the resonance and
start increasing current amplitude. We assume that offset and driving
force are changing proportionally and write~(\ref{eq:nonlinVf}) as
\begin{equation}
\omega_0^2 - \omega^2 + i\delta_i\omega
  + i \delta_0\omega S\left(\frac{|{\bf V}|}{v_0 LB}\right)
= i\omega \frac{{\bf A_0} I}{{\bf V} I_0 - {\bf V_0} I}.
\end{equation}
Note that currents $I$ and $I_0$ here can be just real numbers in
arbitrary units, because multiplying them by
a complex factor will not change the result.

By taking real part of this expression we can find new value of the
resonance frequency (it can change because of other non-linear effects).
From the imaginary part we can find an effective damping
$\delta$:
\begin{equation}\label{eq:delta}
\delta(|{\bf V}|) = \delta_i + \delta_0
S\left(\frac{|{\bf V}|}{v_0 LB}\right)
= \Im\left\{ i \frac{{\bf A_0} I}{{\bf V} I_0 - {\bf V_0} I}\right\}.
\end{equation}

\image{h}{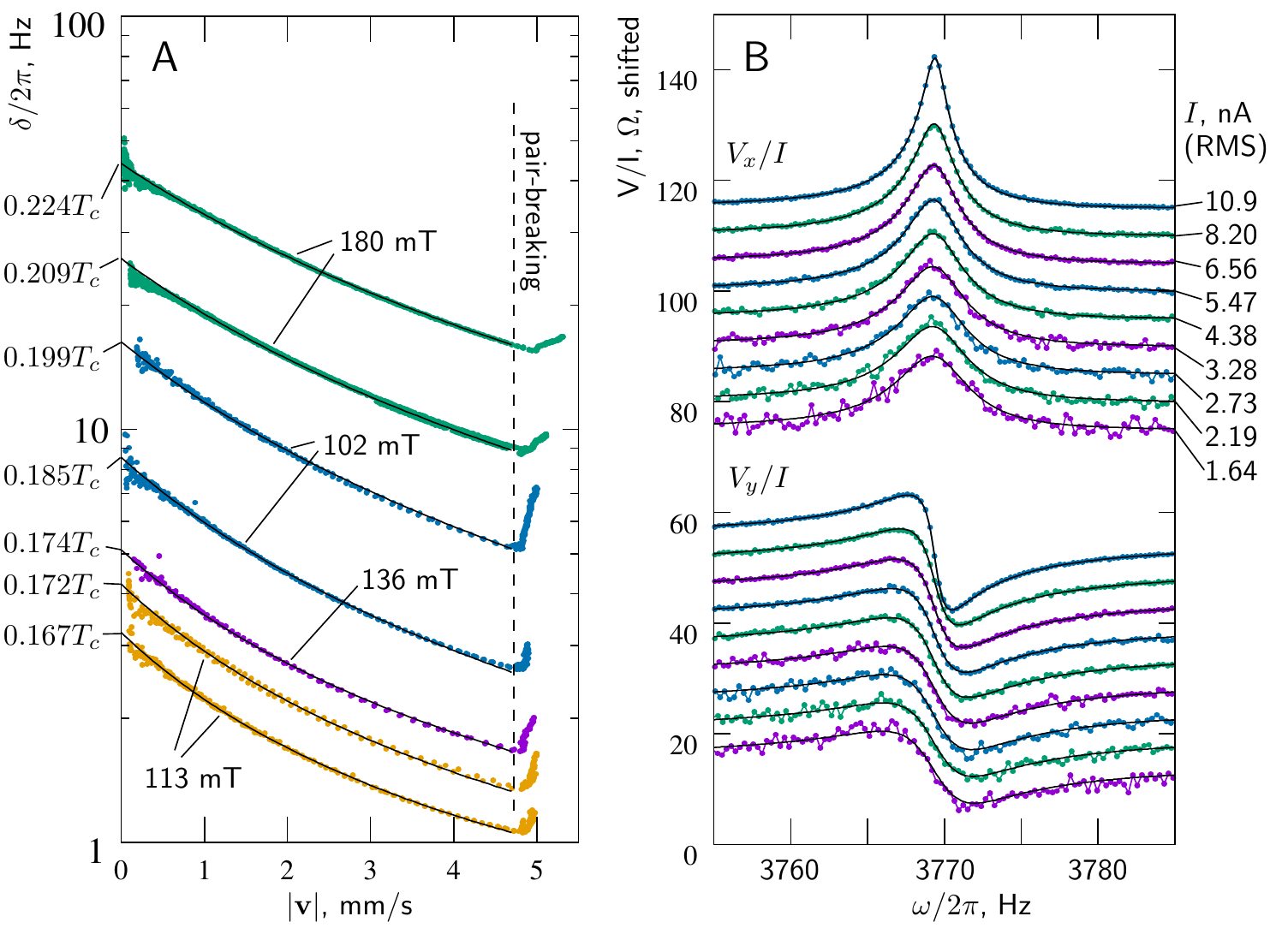}{{\bf A}: measured $\delta(|{\bf v}|)$ at seven different
temperatures and magnetic fields, at zero pressure. Black lines show the
fitting model with same intrinsic damping and same function $S(c)$ for
all data. Temperatures (in $T_c$ units) are shown in assumption that
$s_0=1$. {\bf B}. Frequency sweeps at nine different drive currents are
fitted together with the non-linear equation~(\ref{eq:nonlinVf}) using
intrinsic damping and function $S(c)$ from Fig.~3A, and usual set of free
parameters for Lorentzian fitting: complex amplitude, complex offset,
$\omega_0$ and $\delta(v=0)$}

On Fig.~3A. an example of measured $\delta$ at zero pressure
and seven different temperatures and magnetic fields is shown as a
function of velocity $|{\bf v}|$. A NbTi
wire with $d=4.5\,\mu$m and $L=1.49\,$mm is used. Intrinsic width of NbTi
wire partially originates from motion of vortices in the superconductor and
depends on magnetic field as
\begin{equation}
\delta_i(B) = \delta^i_0 + \delta^i_2 B^2.
\end{equation}
This field dependence can be obtained from measurements of the wire
resonance in vacuum or by using two fitting parameters in $\delta(|{\bf
v}|)$. This is done on Fig.~3A: seven datasets are fitted together using
model with 7+5 parameters: temperature of each measurement, intrinsic
damping $\delta^i_0$ and $\delta^i_2$, and function $S$ described via
parameters $s_0$, $s_1$, $s_2$ introduced in~(\ref{eq:s0s1s2}). Because
$v_0$ and $\delta_0$ have different dependence on
temperature~(\ref{eq:v0delta}) by this kind of fitting it could be
possible to extract temperature without external calibration.
Unfortunately, temperature and $s_0$ have big mutial covariance and
accuracy of this self-calibration for our data is poor. For obtaining
good temperature calibraition it's better to get value of $s_0$ from some
external calibration or use theoretical value $s_0=1$. But regardless of this
choice we will have an accurate model which can be used to extrapolate
$\delta$ to $v=0$ and remove all effects of the non-linear regime.
In all following figures obtaind model will be used in this way.
Temperature calibration can be done as a next independent step.

On Fig.~3B an example of frequency sweeps at nine different drive
currents is shown. All data are fitted together with
equation~(\ref{eq:nonlinVf}) using intrinsic damping and function $S(c)$
obtained from Fig.~3A and usual set of free parameters for Lorentzian fitting:
complex amplitude, complex offset, $\omega_0$ and $\delta(v=0)$.
On Fig.~4A damping for same frequency sweeps is obtained by fitting them
separately with either Lorentzian formula or non-linear equation~(\ref{eq:nonlinVf}).
One can see that even at small currents Lorentzian fit gives
noticeably smaller damping then the "linear" value~$\delta(v=0)$.

\image{h}{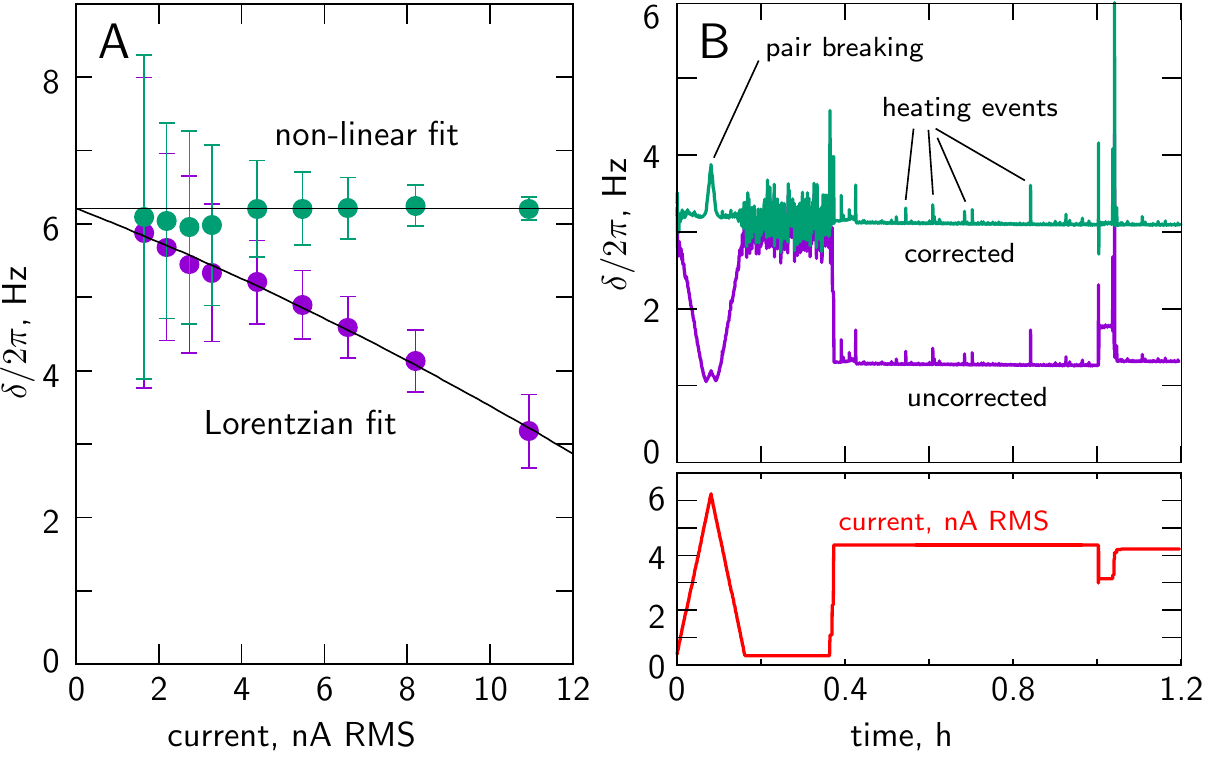}{
{\bf A}: Damping parameter $\delta$ obtained by usual Lorentzian fit of
frequency sweeps from Fig.~3B (lower points); $\delta(v=0)$ found by
fitting same data with the non-linear formula~(\ref{eq:nonlinVf}) (upper
points). Black horizontal line is $\delta(v=0) = 6.213$ found on Fig.~3B
by fitting all sweeps together. {\bf B}: An example of measurement of
$\delta$ as a function of time in "tracking mode". Current (lower plot)
is ramped up and down to reach pair-breaking regime and then kept
constant with a few step adjustments. Measured value of $\delta$ and
corrected value $\delta(v=0)$ are shown on the upper plot. One can see
that corrected value does not depend on drive current (except the
pair-breaking regime) and can be used for temperature measurement. Small
peaks are random heating events in helium caused by natural
radioactivity.}

On Fig.~4B an example of a "tracking mode" measurement is shown. This is
similar to the "amplitude sweep" described above: damping $\delta$ is
extracted from a single measurement of voltage components
using~(\ref{eq:delta}). Normally this is the way of measuring temperature
as a function of time at constant drive, but here a few changes in drive
have been done: in the beginning current was ramped up and down to get
information about non-linear regime and reach pair-braking velocity. Then
current was increased to do measurements in the non-linear regime where
signal-to-noise ratio is better. At the end current have been adjusted
again. On the plot one can see measured value of $\delta$ and corrected
value $\delta(v=0)$. As expected, the corrected value does not depend
on drive current (except in the pair-breaking regime) and can be used for
temperature measurement. Small peaks are random heating events in helium
caused by natural radioactivity.

\subsection*{Conclusion}

This note shows importance of non-linear effects for vibrating-wire
thermometry in superfluid $^3$He-B. Method of removing the non-linearity
is described and tested on experimental data. It is possible to build a simple
model with a few parameters (intrinsic damping and function $S(c)$) which
works at any temperature within ballistic regime (below $0.3 T_c$)
and any magnetic field.

I would like to thank A.Shen, S.Loktev, D.Zmeev and
S.Autti for useful discussions.

\printbibliography

\end{document}